\documentclass[pra,twocolumn,groupedaddress,showpacs,floatfix]{revtex4}

\usepackage{graphics}
\usepackage{graphicx}
\usepackage{bm}
\usepackage{amsmath}
\usepackage{amsfonts} 
\usepackage{amssymb}
\usepackage{latexsym}
\usepackage{color}
\usepackage{mathrsfs}
\usepackage{braket}
\usepackage{enumerate}
\usepackage[usenames,dvipsnames]{xcolor}
\usepackage[colorlinks=true,citecolor=blue,linkcolor=blue,urlcolor=blue,backref=true]{hyperref}
\usepackage{float}
\usepackage{mathtools}
\usepackage[raggedright,tight]{subfigure}  

\begin{document}

\title{Using thermodynamics to identify quantum subsystems€™}
\author{Adam Stokes$^1$}
\author{Prasenjit Deb$^{1,2}$}
\author{Almut Beige$^1$}

\affiliation{$^{1}$The School of Physics and Astronomy, University of Leeds, Leeds LS2 9JT, United Kingdom, The School of Physics and Astronomy} 
\affiliation{$^2$Department of Physics and Center for Astroparticle Physics and Space Science, Bose Institute, Bidhan Nagar Kolkata 700091, India}

\date{\today}

\begin{abstract}
There are many ways to decompose the Hilbert space $\cal H$ of a composite quantum system into tensor product subspaces. Different subsystem decompositions generally imply different interaction Hamiltonians $V$, and therefore different expectation values for subsystem observables. This means that the uniqueness of physical predictions is not guaranteed, despite the uniqueness of the total Hamiltonian $H$ and the total Hilbert space $\cal H$. Here we use Clausius' version of the second law of thermodynamics (CSL) and standard identifications of thermodynamic quantities to identify possible subsystem decompositions. It is shown that agreement with the CSL is obtained, whenever the total Hamiltonian and the subsystem-dependent interaction Hamiltonian commute (i.e. $[H, V ] = 0$). Not imposing this constraint can result in the transfer of heat from a cooler to a hotter subsystem, in conflict with thermodynamics. We also investigate the status of the CSL with respect to non-standard definitions of thermodynamic quantities and quantum subsystems.
\end{abstract}

\maketitle
 
\section{Introduction}\label{1}

More than fifty years ago, it was noted in the context of quantum electrodynamics that there is an inherent relativity present within the definition of a quantum subsystem \cite{power_coulomb_1959,power_nature_1978}. There are many ways of decomposing the total Hilbert space of an atom and its surrounding free radiation field into a tensor product of subspaces. Each decomposition implies a different form of the corresponding atom-field interaction Hamiltonian. The most known forms of this interaction Hamiltonian are the minimal-coupling, the multipolar and the so-called rotating-wave Hamiltonians \cite{drummond_unifying_1987,cohen-tannoudji_photons_1989,baxter_gauge_1990,woolley_gauge_2000,stokes_extending_2012,stokes_gauge_2013}. Although all of these Hamiltonians have the same spectrum and are unitarily equivalent, they result in different predictions for the time evolution of subsystem expectation values. Overcoming this problem requires identifying physically viable subsystem decompositions, a subject that has now spread out to a much wider quantum physics community \cite{nielsen_quantum_2000,zanardi_quantum_2004,barnum_subsystem-independent_2004,dugic_what_2006}. 

Suppose we decompose the total Hilbert space ${\mathcal H}$ of two interacting quantum systems $a$ and $b$ in two different ways, 
\begin{eqnarray}
{\cal H} = {\cal H}_a \otimes {\cal H}_b = {\cal H'}_a \otimes {\cal H'}_b \, . 
\end{eqnarray}
Moreover, we introduce the orthonormal bases $\{\ket{i_c}\}$ of ${\cal H}_c$ and the orthonormal bases 
$\{\ket{i'_c}\}$ of ${\cal H'}_c$ with $c=a,b$ respectively. Both sets of product states, $\{\ket{i_a,j_b}\}$ and $\{\ket{i'_a,j'_b}\}$, form an orthonormal basis in ${\cal H}$, so they are related by a unitary transformation $U$;
\begin{eqnarray}
\ket{i'_a,j'_b}=U\ket{i_a,j_b} \, . 
\end{eqnarray}
Consequently, the operator space ${\cal L}({\cal H})$, consisting of the linear maps on $\cal H$, possesses bases $\{O_a^\mu\otimes O_b^\nu \}$ and $\{O_a^{'\mu}\otimes O_b^{'\nu}\}$, such that 
\begin{eqnarray}
O_a^{'\mu}\otimes O_b^{'\nu} = U O_a^\mu \otimes O_b^\nu U^\dagger \, . 
\end{eqnarray}
Subsystem relativity arises when the unitary transformation $U$ is not of the form $U_a\otimes U_b$. In this case the tensor product decompositions ${\cal H}_a\otimes {\cal H}_b$ and ${\cal H'}_a\otimes {\cal H'}_b$ define physically distinct subsystems $a$ and $b$. Adopting the decomposition ${\cal H} = {\cal H}_a\otimes {\cal H}_b$, observables associated with subsystem $a$ have the form
\begin{align}
{\mathcal O}_a = \sum_\mu c_\mu O_a^\mu\otimes I_b \, ,
\end{align}
where $I$ denotes an identity operator. These operators do not act on ${\cal H}_b$ but generally have a non-trivial effect in ${\mathcal H'}_b$. On the other hand, adopting the decomposition ${\cal H} = {\cal H'}_a\otimes {\cal H'}_b$, observables associated with subsystem $a$ have the form
\begin{align}
{\mathcal O}'_a = \sum_\mu c'_\mu O_a^{'\mu}\otimes I'_b \, .
\end{align}
These operators do not act on ${\cal H'}_b$ but generally have a non-trivial effect in ${\mathcal H}_b$.

In general, the total Hamiltonian $H$ of a composite quantum system can be partitioned into free and interaction components in many different ways. Each different partition corresponds to a different subsystem decomposition. For example
\begin{align}\label{h2}
& H\{O_a^\mu,O_b^\nu\} \nonumber \\ &:=H_a\{O_a^\mu\} \otimes I_b + I_a \otimes H_b\{O_b^\nu\} + V\{O_a^\mu \otimes O_b^\nu\}\nonumber \\ &= H_a\{O_a^{'\mu}\} \otimes I'_b + I'_a \otimes H_b\{O_b^{'\nu}\} + V'\{O_a^{'\mu}\otimes O_b^{'\nu}\} \nonumber \\ &=: H'\{O_a^{'\mu},O_b^{'\nu}\} \,
\end{align}
where $H_a$, $H_b$, $V$, $V'$, $H$ and $H'$ denote specific functional forms of basis operators. In terms of the distinct operator bases associated with different subsystem decompositions, the total Hamiltonian $H$ possesses different interaction components and has different functional forms. Moreover, we have
\begin{align}
&\langle {\mathcal O}_a \rangle_t \nonumber \\ &= \bra{i_a,j_b}e^{iH\{O_a^\mu,O_b^\nu\}t}{\cal O}_a e^{-iH\{O_a^\mu,O_b^\nu\}t} \ket{i_a,j_b}\nonumber \\ &\neq  \bra{i'_a,j'_b}e^{iH'\{O_a^{'\mu},O_b^{'\nu}\}t}{\cal O}'_a e^{-iH'\{O_a^{'\mu},O_b^{'\nu}\}t} \ket{i'_a,j'_b} \nonumber \\&= \langle {\cal O}'_a \rangle_t \, ,
\end{align}
which means that the time evolution of subsystem observables depends in general on the chosen subsystem decomposition.

Naturally, the fundamentally important question occurs as to which of the above predictions correctly describes the physical subsystem $a$. Ultimately this question must be answered by comparison with empirical findings. For example, using standard open quantum systems approaches, it has been shown that non-negligible concentrations of bare energy, 
\begin{eqnarray} \label{H0}
H_0 &=& H_a+H_b \, ,
\end{eqnarray}
can occur, when an atom couples to an electromagnetic field \cite{kurcz_energy_2010,stokes_extending_2012}. Rejecting this bare energy build-up as non-physical (it is not something observed empirically) restricts the form of the atom-field interaction to one which conserves the bare energy. Such a restriction may be especially nontrivial given that it may seem incompatible with the form of interaction obtained \emph{a fortiori} from elementary first principles. 

The laws of thermodynamics are some of the oldest and best-established laws of modern physics and are extensively supported by centuries of empirical evidence. Here we propose applying Clausius' form of the second law of thermodynamics (CSL) in quantum physics to restrict the possible forms of quantum interactions and subsystem decompositions of composite quantum systems. The CSL \cite{clausius_ueber_1854} can be stated as follows:
\begin{center}
\begin{minipage}{7cm}
\item \textit{No process is possible whose sole result is the transfer of heat from a cooler to a hotter body.}
\end{minipage}
\end{center} 
Since the laws of thermodynamics are usually taken to refer to average quantities pertaining to large numbers of elementary systems, their status within the microscopic domain, which is generally believed to require a quantum-theoretic modelling, is less clear. In the quantum setting, studies of the thermodynamical laws have largely taken place within the context of open-quantum system's theory, wherein small systems are coupled to thermal reservoirs with infinite degrees of freedom. General quantum versions of basic thermodynamic results have been established in this context \cite{breuer_theory_2007,alicki_quantum_2007,kosloff_quantum_2013}. Among these are the fluctuation theorem-type results which suggest that the statistical laws of thermodynamics may occasionally be transiently violated for individual systems \cite{evans_probability_1993,evans_equilibrium_1994}. There are also experimental results which support this possibility \cite{wang_experimental_2002,gieseler_dynamic_2014}.

Situations consisting of only a small number of interacting quantum systems have received widespread attention only relatively recently. Most of the work on quantum thermodynamics focusses on identifying quantum versions of the thermodynamic quantities and their associated laws for a single elementary system in contact with a heat bath \cite{esposito_entropy_2010,esposito_second_2011,horodecki_fundamental_2013,brandao_second_2015}. Our focus here is on the Clausius form of the second law for interacting elementary systems initially in contact with separate heat baths so as to be prepared in thermal states. These are then isolated and allowed to evolve. Results relating to this situation are fewer and further between. Tasaki has shown that a general entropic version of the second law of thermodynamics can be proven in this situation \cite{tasaki_jarzynski_2000}. Weimer \emph{et al.}~have developed an energy-flux formalism based on the LEMBAS (local measurement basis) principle, which they use to give general definitions of heat and work within the interacting setting \cite{weimer_local_2008}. This formalism is further studied and extended to include open quantum systems in \cite{hossein-nejad_work_2015}, where it is also shown to give rise to an entropic version of the second law. Esposito {\em et al.}~have given general identifications of heat and work for a system in contact with several heat baths, and have also identified a useful partition of the change in the system's von Neumann entropy into an always positive entropy production term and a reversible heat-flux term \cite{esposito_entropy_2010,esposito_second_2011}.

In this paper we show that the result of Tasaki can be extended to include Clausius' version of the second law of thermodynamics, but only provided one assumes that the bare energy is conserved. This condition can be expressed through the vanishing of any one of three equivalent commutators as
\begin{align}\label{result2}
[H_0,V] \equiv [H,V] \equiv [H_0,H] = 0
\end{align}
where $H=H_0+V$. Whether or not condition (\ref{result2}) holds depends on the particular interaction, i.e., subsystem decomposition, selected within a model. Assuming the standard identifications of heat and quantum subsystems are correct, then for the CSL to hold it is sufficient that condition (\ref{result2}) holds. Rejecting condition (\ref{result2}) means accepting at least transient violations of the CSL, or employing non-standard definitions of heat and/or quantum subsystems.

There are five sections in this paper. In section \ref{theory} we recap the standard quantum thermodynamics definitions of heat, work and entropy. We then determine the condition (\ref{result2}) as sufficient in order that the CSL holds for two interacting systems initially prepared in thermal states at different temperatures. We introduce a particularly simple interacting oscillator system which we use to investigate the CSL in the following sections. In section \ref{vio} we demonstrate violations of the CSL using a particular interaction Hamiltonian for the simple two-oscillator system introduced in section \ref{theory}. In section \ref{4} we extend our previous analyses by adopting alternative identifications of heat, work and entropy \cite{weimer_local_2008,hossein-nejad_work_2015, esposito_entropy_2010}. Finally we summarise our findings in section \ref{conc}.
 
\section{Theoretical background}\label{theory}

In this section we begin by reviewing the standard definitions of the basic thermodynamic quantities. We then define the heat transfer between interacting subsystems. Finally we determine a sufficient condition (\ref{result2}) in order that the CSL holds for the interacting system.

\subsection{Energy, heat and entropy: the standard definitions}\label{standard}

The first law of thermodynamics states that
\begin{align}\label{1st}
{\Delta H}={\Delta  W}+{\Delta Q}
\end{align}
where $H$ is a systems total energy, $\Delta W$ is the change in the external work energy and $\Delta Q$ is the change in heat energy. In quantum theory the expected rate of change of the total energy is given by
\begin{align}\label{heat}
\left\langle {dH \over dt} \right\rangle = {\rm tr}({\dot H}\rho) + {\rm tr}(H{\dot \rho})
\end{align}
where $\rho$ is the system's density matrix. The first term on the right-hand side in (\ref{heat}) can be identified as the work contribution $d\langle W \rangle/dt$ and the second term can be identified as the heat contribution $d\langle Q\rangle/dt$ \cite{alicki_quantum_1979,kosloff_beyond_1984,kieu_second_2004,henrich_small_2006,henrich_driven_2007}. When the total energy is not explicitly time-dependent the work contribution vanishes so that $\Delta H = \Delta Q$. If we identify the entropy via the Gibbs relation
\begin{align}\label{gibbs}
\Delta S = \beta \Delta Q
\end{align}
where $\beta$ is the inverse temperature, then we see that $\Delta S = 0$ if and only if $\Delta H = \Delta W$. In this paper we adopt the above definitions of work, heat and entropy, and apply them on the level of two quantum subsystems $a$ and $b$ initially prepared in thermal states. The corresponding subsystem quantities are respectively denoted $W_c$, $Q_c$ and $S_c$ with $c=a,b$. In the following, we consider the general case in which the subsystems are interacting, but in which the energies $H_a$, $H_b$ and $H$ are not explicitly time-dependent.

For concreteness and simplicity we consider for the most part two interacting quantum harmonic oscillators (QHOs) labelled $a$ and $b$, though our main results hold for arbitrary interacting quantum systems. We consider the situation in which the oscillators $a$ and $b$ are prepared at $t=0$ in the thermal states 
\begin{align}\label{rho0}
\rho_c^{\rm th.} =  {1\over Z_c}e^{-\beta_cH_c}
\end{align}
with $c=a,b$. Here $\beta_c$ is the inverse temperature of oscillator $c$ and the $H_c$'s are the oscillator energy operators defined by
\begin{align}\label{Hab}
H_a = \omega_a a^\dagger a, \qquad H_b = \omega_b b^\dagger b.
\end{align}
The operators $a$ and $b$ in this equation are bosonic annihilation operators satisfying $[a,a^\dagger]=1=[b,b^\dagger]$ and which act respectively, within the abstract seperable Hilbert spaces ${\mathcal H}_a$ and ${\mathcal H}_b$ of the corresponding oscillators. The eigenstates of $H_0=H_a+H_b$ form a basis of orthonormal states in the composite system's Hilbert space ${\mathcal H}={\mathcal H}_a\otimes {\mathcal H}_b$. We denote these states $\{\ket{n_a,n_b} =\ket{n_a}\otimes \ket{n_b}\}$, and refer to them as the bare (as oppossed to dressed) eigenstates. The energy eigenvalue corresponding to the state $\ket{n_c}$ with $c=a,b$ is denoted $\omega_c^n =n_c\omega_c$. The functions $Z_c = {\rm tr}(e^{-\beta_cH_c})$ are included in (\ref{rho0}) to ensure the correct normalisation of the thermal states. Starting with the initial state of the composite $ab$ system \begin{align}\label{rho02}
\rho= \rho_a^{\rm th.}\otimes \rho_b^{\rm th.}
\end{align}
enables one to form clear questions about heat transfer between the two systems at subsequent times.

\subsection{Heat transfer according to the standard definitions}

The identification of the heat changes $\Delta Q_a$ and $\Delta Q_b$ of the two subsystems within a time $t>0$ follows from the standard definitions given above in Section \ref{standard}. Since the Hamiltonian is time-independent there are no work contributions. So the heat energies of the subsystems are given by the bare energies $H_a$ and $H_b$. In what follows the basic quantity we will consider is the heat transfer from oscillator $a$ into oscillator $b$ within a time $t\geq 0$. We denote this quantity $\Delta Q_{a\to b}$ and define it as follows
\begin{align}
&\Delta Q_{a\to b} := \Delta Q_b - \Delta Q_a, \nonumber \\ 
&\Delta Q_c = \langle H_c(t) - H_c(0)\rangle_{\rho}
\end{align}
with $c=a,b$. Here $\langle \cdot \rangle_{\rho}$ denotes the standard quantum expectation value taken in the state $\rho$ given in (\ref{rho02}). In terms of $\Delta Q_{a\to b}$ the CSL can be expressed simply as
\begin{align}\label{CSL}
{\rm sgn}\Delta  Q_{a\to b} = {\rm sgn}(\beta_b-\beta_a)
\end{align}
which in words, states that the (heat) energy acquired by oscillator $b$ from oscillator $a$, within a time $t$, is positive whenever oscillator $a$ is initially hotter, and is negative whenever $b$ is initially hotter.

\subsection{Conservation of bare energy and the second law}\label{con}

We will now show that the condition (\ref{result2}) is a sufficient condition in order that the CSL holds. The proof in this section builds upon the work of Tasaki \cite{tasaki_jarzynski_2000}, which uses the techniques of Jarzynski \cite{jarzynski_nonequilibrium_1997}. We begin by defining the free entropy change
\begin{align}\label{S0}
\Delta S_0 &:= \beta_a \langle H_{a}(t) - H_{a}(0)\rangle_{\rho}+\beta_b \langle H_{b}(t) - H_{b}(0)\rangle_{\rho}\nonumber \\ &=\beta_a\Delta Q_a +\beta_b \Delta Q_b
\end{align}
where again the initial state $\rho$ is given in (\ref{rho02}). Next we define the ``classical" average of a given function $f:{\mathbb R}^4\to {\mathbb R}$ as
\begin{align}\label{classav}
{\rm E}[f]_t &:= \nonumber \\& \sum_{nmpq} &{e^{-\beta_a \omega_a^n}e^{-\beta_b \omega_b^m}\over Z_a Z_b}|U_{pq;nm}(t)|^2 f(\omega^n_a,\omega^m_b,\omega^p_a,\omega^q_b)
\end{align}
where $U_{pq;nm}(t) = \bra{p_a,q_b}U(t)\ket{n_a,m_b}$ are the transition amplitudes between the bare states.

It is straightforward to verify that the four projection functions defined by $\omega_a(w,x,y,z) = w$, $\omega_b(w,x,y,z) = x$, $\omega'_a(w,x,y,z) = y$ and $\omega'_b(w,x,y,z) = z$ satisfy
\begin{eqnarray}\label{E0}
{\rm E}[\omega_a]_t &=& \sum_{nmpq} {e^{-\beta_a \omega_a^n}e^{-\beta_b \omega_b^m}\over Z_a Z_b} \omega^{n}_a \nonumber \\ && \times \bra{n_a,m_b}U^{-1}(t)\ket{p_a,q_b}\bra{p_a,q_b}U(t)\ket{n_a,m_b} \nonumber \\ 
&=& \sum_{nm}\omega_a^n {e^{-\beta_a \omega_a^n}e^{-\beta_b \omega_b^m}\over Z_a Z_b} \nonumber \\ 
&=& {\rm tr}(H_a(0)\rho)  = \langle H_{a}(0)\rangle_\rho \, , \nonumber \\
{\rm E}[\omega_b]_t &=& \sum_{nmpq} {e^{-\beta_a \omega_a^n}e^{-\beta_b \omega_b^m}\over Z_a Z_b} \omega^m_b \nonumber \\ 
&& \times \bra{n_a,m_b}U^{-1}(t)\ket{p_a,q_b}\bra{p_a,q_b}U(t)\ket{n_a,m_b}  \nonumber \\ 
&=& \sum_{nm}\omega_b^m {e^{-\beta_a \omega_a^n}e^{-\beta_b \omega_b^m}\over Z_a Z_b} \nonumber \\ 
&=& {\rm tr}(H_b(0)\rho) = \langle H_b(0)\rangle_\rho  
\end{eqnarray}
where we have used the completeness of the bare states. Similarly
\begin{align}\label{Et}
&{\rm E}[\omega'_a]_t = \sum_{nm} {e^{-\beta_a \omega_a^n}e^{-\beta_b \omega_b^m}\over Z_a Z_b} \nonumber \\  \times & \bra{n_a,m_b} U^{-1}(t) \left[\sum_{pq}  \omega_a^p \ket{p_a,q_b}\bra{p_a,q_b}\right] U(t) \ket{n_a,m_b}  \nonumber \\ & = {\rm tr}(H_a(t)\rho) = \langle H_a(t) \rangle_\rho \, , \nonumber \\
&{\rm E}[\omega'_b]_t = \sum_{nm} {e^{-\beta_a \omega_a^n}e^{-\beta_b \omega_b^m}\over Z_a Z_b} \nonumber \\  \times & \bra{n_a,m_b} U^{-1}(t) \left[\sum_{pq}  \omega_b^q \ket{p_a,q_b}\bra{p_a,q_b}\right] U(t) \ket{n_a,m_b}  \nonumber \\ & = {\rm tr}(H_b(t)\rho) = \langle H_b(t) \rangle_\rho 
\end{align}
where the operators in square brackets are simply the spectral representations of $H_a(0)$ and $H_b(0)$. Furthermore, it follows from (\ref{classav}) that
\begin{align}\label{E1}
&{\rm E}[e^{\beta_a(\omega_a-\omega'_a)+\beta_b(\omega_b-\omega'_b)}]_t =  \sum_{pq;nm} {e^{-\beta_a \omega_a^p}e^{-\beta_b \omega_b^q}\over Z_a Z_b} \nonumber \\ & \times \bra{p_a,q_b}U(t)\ket{n_a,m_b} \bra{n_a,m_b}U^{-1}(t)\ket{p_a,q_b} \nonumber \\ &= \sum_{pq} {e^{-\beta_a \omega_a^p}e^{-\beta_b \omega_b^q}\over Z_a Z_b}=1
\end{align}
where again we have used the completeness of the bare states. From (\ref{E0}), (\ref{Et}), (\ref{E1}) and the Jensen inequality $ e^{{\rm E}[f]_t} \leq {\rm E}[e^f]_t $, it now follows that $e^{-\Delta S_0} \leq 1$ and hence
\begin{align}\label{ESL}
\Delta S_0 \geq 0.
\end{align}
This result can be viewed as an entropic version of the second law of thermodynamics (ESL) pertaining to the quantum system. The existence of this ESL supports the identifications of heat and entropy initially made in equation (\ref{S0}). We remark that these identifications have also been employed elsewhere \cite{kurcz_rotating_2010,blasone_quantum_2011}. It is therefore of interest to ask whether or not the CSL also holds when these identifications are made.

In order to go further and make contact with the CSL, we must resort to our assumption (\ref{result2}), which implies
\begin{align}
\Delta Q_a = -\Delta Q_b,\qquad \Delta Q_{a\to b} = 2\Delta Q_b.
\end{align}
From this it follows using (\ref{ESL}) that
\begin{align}
(\beta_b -\beta_a)\Delta Q_{a\to b} \geq 0,
\end{align}
which is nothing but the CSL (\ref{CSL}).

Note that in the present context the CSL is a significantly stronger statement than the ESL, because it requires the additional nontrivial assumption (\ref{result2}). While the ESL (\ref{ESL}) holds quite generally for the initial thermal state (\ref{rho02}) regardless of the form of $V$, the CSL (\ref{CSL}) will not generally hold when condition (\ref{result2}) is not met.

\subsection{General form of heat transfer}

In order to obtain a general expression for the heat transfer $\Delta Q_{a\to b}$ we assume that the operators $a(t)$ and $b(t)$ can be expanded as the following linear combinations of operators at $t=0$
\begin{align}\label{abgen}
&a(t)=f_a(t) a + g_a(t) a^\dagger + f_b(t) b + g_b(t) b^\dagger \nonumber \\
&b(t)=p_a(t) a + q_a(t) a^\dagger + p_b(t) b + q_b(t) b^\dagger
\end{align}
where $a =a(0)$ and $b=b(0)$. Such expansions result whenever the interaction Hamiltonian $V$ contains only linear and quadratic terms in the operators $a,b,a^\dagger,b^\dagger$. This includes almost all interactions of practical interest. A straightforward calculation now yields
\begin{align}\label{Qab}
\Delta Q_a =& \omega_a[(|f_a|^2+|g_a|^2-1)X_a \nonumber \\ & + (|f_b|^2+|g_b|^2)X_b +|g_a|^2 + |g_b|^2]\nonumber \\ 
\Delta Q_b =& \omega_b[(|p_b|^2+|q_b|^2-1)X_b \nonumber \\ & + (|p_a|^2+|q_a|^2)X_a +|q_a|^2 + |q_b|^2] \, .
\end{align}
Here
\begin{align}\label{X}
X_c = {1\over e^{\beta_c \omega_c} - 1} 
\end{align}
with $c=a,b$. The calculation of $\Delta Q_{a\to b}$ now reduces to the determination of the time-dependent functions in (\ref{abgen}) for a given choice of interaction Hamiltonian $V$.

\section{Violations of the CSL}\label{vio}

In this section we consider the case of two interacting oscillators, and demonstrate violations of the CSL for a particular choice of interaction Hamiltonian. 

\subsection{Forms of interaction for which the CSL holds}

We have established that whenever condition (\ref{result2}) is met the CSL will hold. In many areas of quantum theory, such as quantum optics, interactions $V$ satisfying this condition are usually obtained through approximations of Hamiltonians derived from first principles. The most obvious example is the famous Jaynes-Cummings model, which has become the paradigm of solvable light-matter systems. The Jaynes-Cummings interaction results as a so-called rotating-wave approximation (RWA) of a more fundamental Hamiltonian for which condition (\ref{result2}) does not hold. The RWA eliminates terms in the interaction which do not conserve the bare energy $H_0$.

\begin{figure}[h]
\begin{minipage}{\columnwidth}
\begin{center}
\hspace*{-2mm}\includegraphics[scale=0.7]{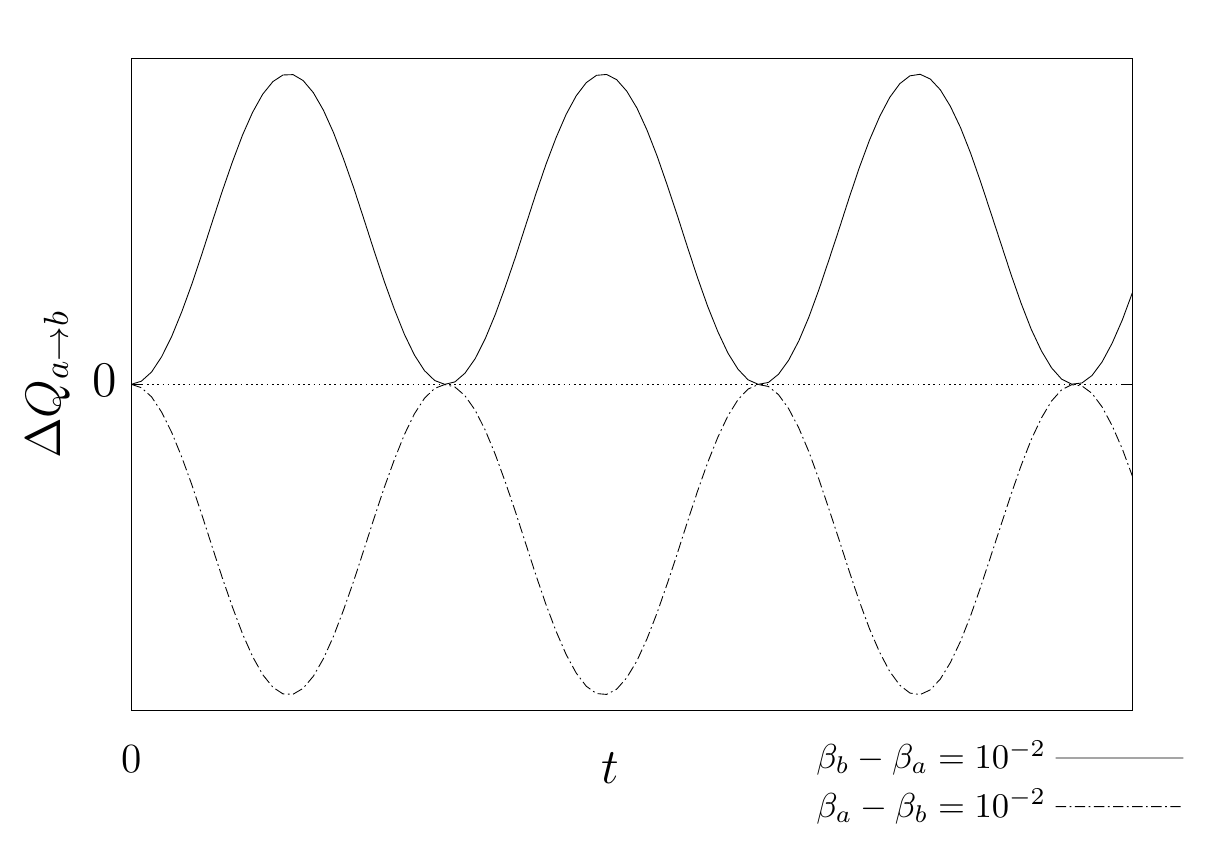}
\end{center}
\vspace*{-5mm}\caption{Plot of the rotating-wave approximated heat transfer $\Delta Q_{a\to b}(t)$ with $g/\omega= 1/10$. The solid curve corresponds to the case $T_a-T_b =50K$, while the dashed curve corresponds to the case $T_b-T_a =50K$. Since the bare energy $H_0$ is conserved the heat gained by one oscillator must equal the heat lost by the other. The curves are therefore mirrored in the time axis. In accordance with the CSL, the solid curve is positive semi-definite for all $t$, and the dashed curve is negative semi-definite for all $t$.} \label{f1}
\end{minipage}
\end{figure}

In the context of the interacting oscillator system we are considering the RWA could for example take the form
\begin{align}\label{rwa}
V = ig(a^\dagger+a)\otimes (b^\dagger -b) \longrightarrow ig(a\otimes b^\dagger - a^\dagger\otimes b)
\end{align}
where $g$ is a real coupling parameter. Note that before the RWA $[H_0,V]\neq 0$, while afterwards $[H_0,V]=0$. Within this RWA the dynamics of the interacting system are especially simple to solve. Assuming for further simplicity the resonance condition $\omega_a=\omega= \omega_b$ we obtain the solution
\begin{align}
a(t) &= e^{-i\omega t} [a \cos gt -b\sin gt], \nonumber \\
b(t) &= e^{-i\omega t} [b \cos gt +a\sin gt].
\end{align}
Comparison with (\ref{abgen}) shows that for this choice of interaction
\begin{align}
f_a =  e^{-i\omega t}\cos gt = p_b,\qquad
f_b = -e^{-i\omega t} \sin gt = -p_a,
\end{align}
while $g_a=g_b=0=q_a = q_b$. Substitution of these expressions into (\ref{Qab}) then yields
\begin{align}
\Delta Q_{a\to b} = 2\omega(X_a-X_b)\sin^2 gt.
\end{align}
Since according to (\ref{X}) we have that ${\rm sgn}(X_a-X_b) = {\rm sgn}(\beta_b-\beta_a)$ we have verified explicitly that the CSL holds for the rotating-wave approximated interaction. Figure \ref{f1} illustrates the behavior of $\Delta Q_{a\to b}$, and clearly shows that it is quite in-line with one's intuition regarding the CSL.


\subsection{Violations of the CSL}\label{violatess}

There are numerous interactions one might consider in order to exhibit a violation of the CSL. Indeed, any of the more fundamental interactions to which the RWA is usually applied do not have the property $[H,V]=0$. They therefore open the door to violations of the CSL. Examples of such interactions include minimal coupling of oscillators and linear-coupling of oscillators. The former of these is the more physical of the two, as the minimal coupling prescription is a major ingredient in elementary theories that are built from first principles. However, because it contains additional self energy terms (terms quadratic in $a^\dagger$ for example) its associated dynamics are more difficult to solve (though still quite possible).

For simplicity we consider the straightforward approach of avoiding the RWA as prescribed in (\ref{rwa}) thereby adopting the linear interaction
\begin{align}\label{v}
V = ig(a^\dagger + a)\otimes (b^\dagger-b).
\end{align}
Again, for simplicity of solution, we retain the resonance condition $\omega_a=\omega=\omega_b$. This is despite the fact that considering off-resonant oscillators actually allows one to exhibit additional violations of the CSL.

When on resonance the introduction of new modes $c_+=(a+ib)/\sqrt{2}$ and $c_-=(b+ia)/\sqrt{2}$ decouples the equations of motion, which can then be solved simply giving
\begin{align}
c_\pm(t) =&c_\pm \left[\cos\mu_\pm t - {i(\omega\pm g)\over \mu_\pm} \sin\mu_\pm t\right]+c_\pm^\dagger{ig\over \mu_\pm}\sin\mu_\pm t
\end{align}
where $\mu_{\pm} = \sqrt{\omega^2 \pm 2\omega g}$. From these solutions the solutions for $a(t)$ and $b(t)$ are easily found and the time-dependent functions in (\ref{abgen}) can then be read-off. Subsequent substitution of these functions into (\ref{Qab}) then yields expressions for $\Delta Q_a, \Delta Q_b$ and $\Delta Q_{a\to b}$.

\begin{figure}[h]
\begin{minipage}{\columnwidth}
\begin{center}
\hspace*{-6mm}\includegraphics[scale=0.7]{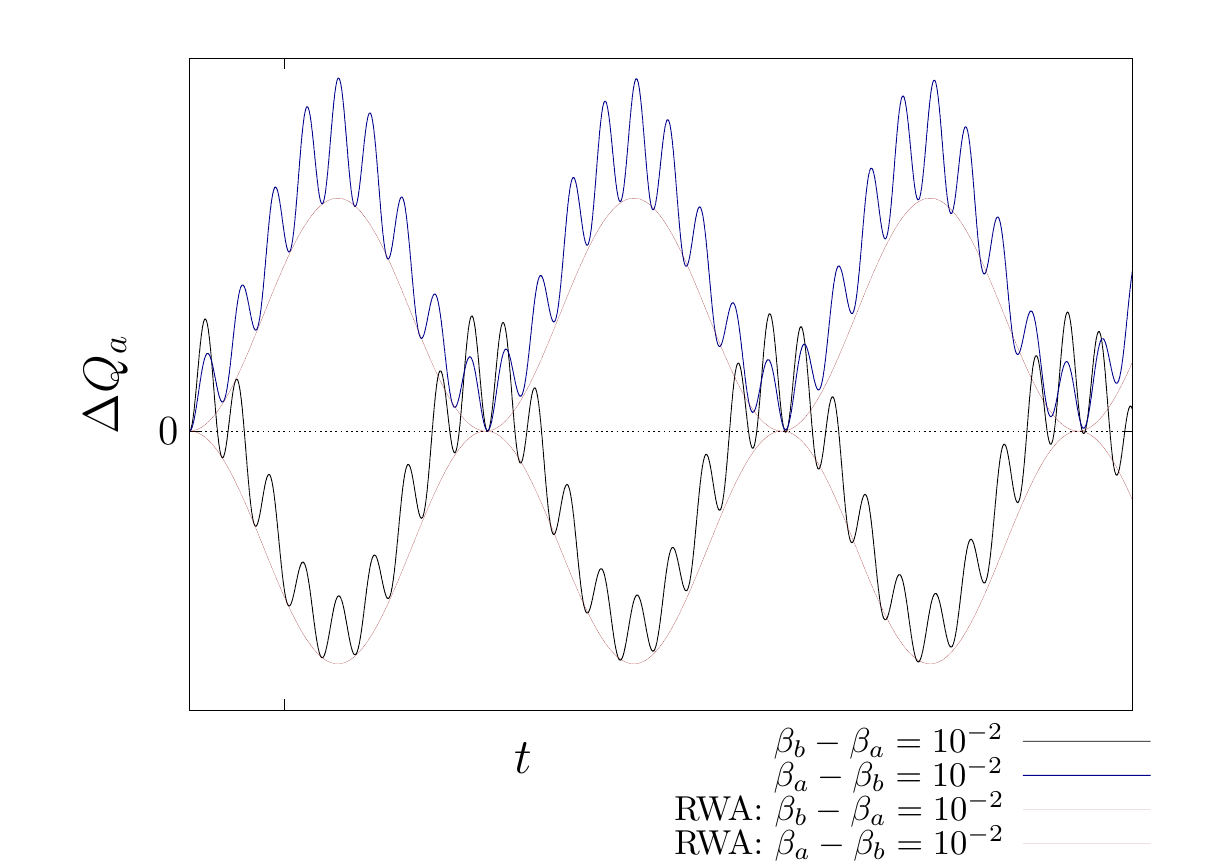}
\end{center}
\vspace*{-5mm}\caption{Plot of $\Delta Q_a(t)$ inside and outside the RWA at fixed temperature difference $|T_a-T_b| =50K$ with $g/\omega=1/10$. Because $g$ is sufficiently large, deviations from the RWA result are encountered as expected. Since the bare energy $H_0$ is not conserved outside the RWA the two corresponding curves are not mirrored in the time axis, unlike the RWA curves. Outside the RWA oscillator $a$ absorbs heat when it is both initially cooler and initially hotter than oscillator $b$.} \label{f2}
\end{minipage}
\end{figure}

Figure \ref{f2} shows how the behavior in $\Delta Q_a$ can differ in an essential way when $[H,V]\neq 0$. Specifically, it is clear that $\Delta Q_a \neq -\Delta Q_b$, and that actually the hotter (as well as the cooler) oscillator begins to absorb heat (bare energy). This behavior by itself does not necessarily imply a violation of the CSL, because when $H_0$ is not conserved one cannot necessarily interpret the heat absorbed by the hotter oscillator as having been transferred from the cooler oscillator. Put differently, we have proved that ${\dot H}_0=0$ is a sufficient condition in order that the CSL holds, but we have not proved that it is also a necessary condition.

\begin{figure}[h]
\begin{minipage}{\columnwidth}
\begin{center}
\hspace*{-2mm}\includegraphics[scale=0.7]{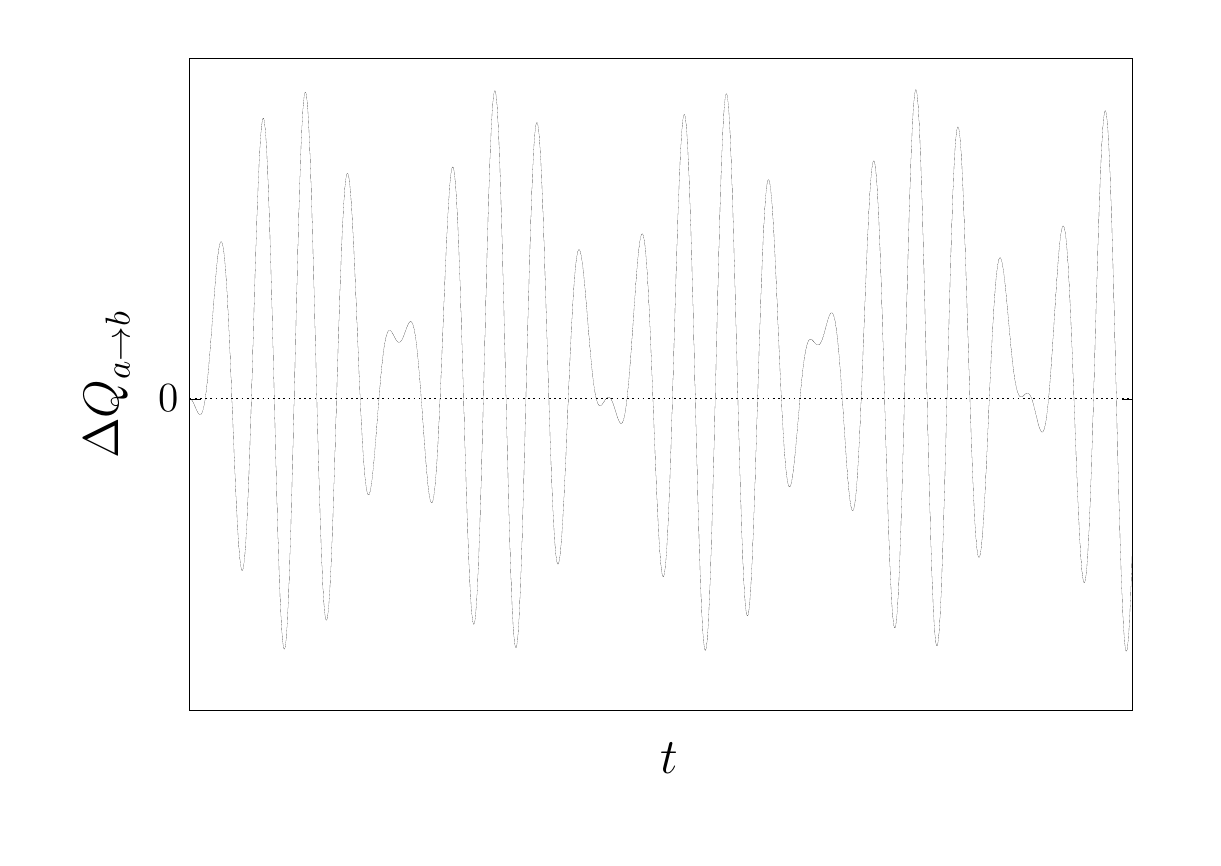}
\end{center}
\vspace*{-5mm}\caption{Plot of $\Delta Q_{a\to b}(t)$ with $g/\omega \approx 1/2$, and with $\beta_b>\beta_a$. The heat transfer exhibits large oscillations in time, resulting in positive and negative values, and hence in a violation of the CSL.} \label{f3}
\end{minipage}
\end{figure}

To determine whether the CSL is truly violated the relevant quantity is $\Delta Q_{a\to b}$. Figure \ref{f3} shows how $\Delta Q_{a\to b}$ oscillates with $t$ and becomes negative even when $\beta_b>\beta_a$. In fact $\Delta Q_{a\to b}$ also oscillates when $t$ is fixed but $g$ varies. A critical value occurs at $g=\omega/2$, at which point $\mu_- =0$ making the quantities of interest singular.

\begin{figure}[h]
\begin{minipage}{\columnwidth}
\begin{center}
\hspace*{-4mm}\includegraphics[scale=0.7]{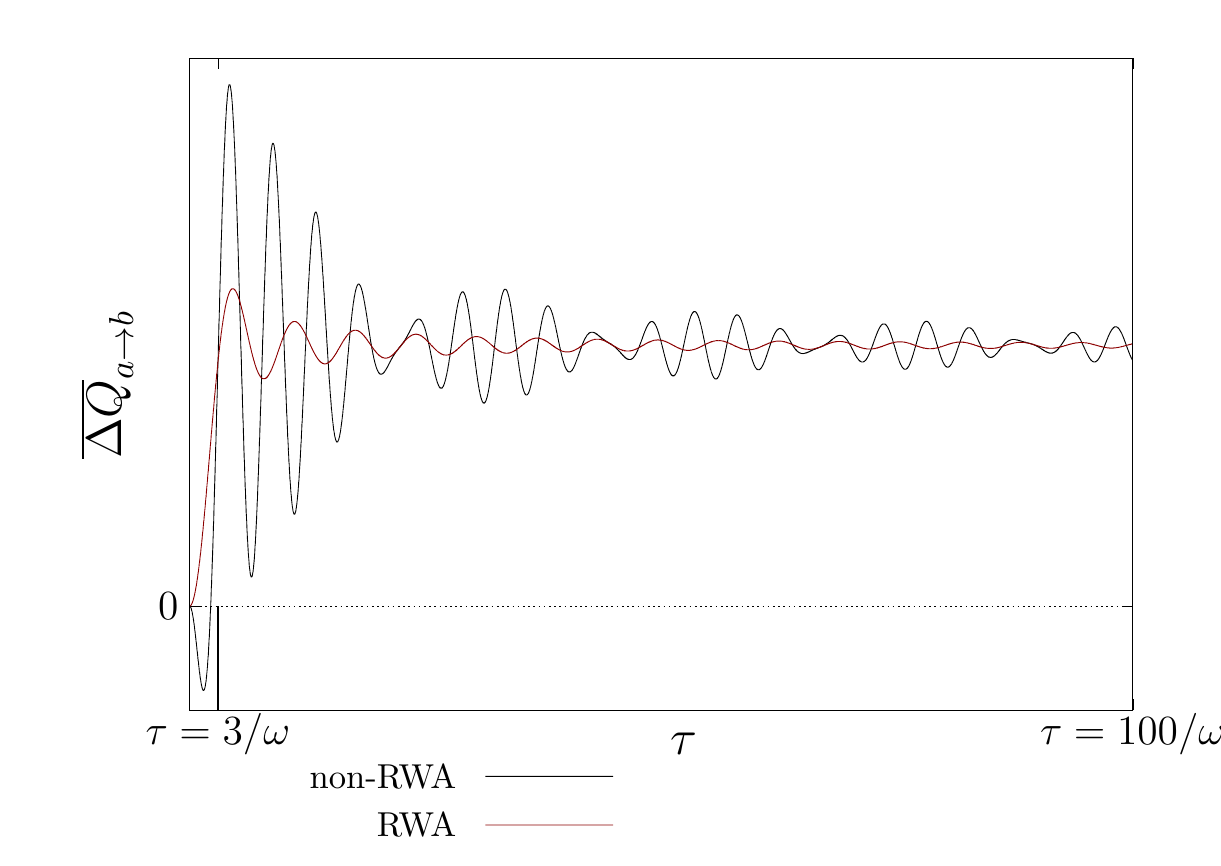}
\end{center}
\vspace*{-5mm}\caption{Plot of $\overline{\Delta Q}_a(\tau)$ inside and outside the RWA at fixed temperature difference $T_a-T_b =50K$, i.e., $\beta_b-\beta_a = 10^{-2}$, with $g/\omega =0.49\omega$. The time averged heat transfer is positive for $\tau$ larger than a few oscillator cycles as indicated by the vertical line at $\tau=3/\omega$. Within the weak coupling regime, the averaged heat transfer is negative (which indicates a violation of the CSL) only transiently for small values of $\tau$.} \label{f4}
\end{minipage}
\end{figure}

The amplitudes of the oscillations in $\Delta Q_{a\to b}$ begin to rapidly diverge as the value of $g$ surpasses $\omega/2$, a regime which can be classed as the strong coupling regime. However, for sufficiently weak coupling, $g<\omega/2$, the non-divergent oscillatory nature of $\Delta Q_{a\to b}$ suggests an interpretation of the violations of the CSL as transient phenomena. This interpretation is bolstered by the behavior of the time-averaged heat transfer
\begin{align}
\overline{\Delta Q}_{a\to b}(\tau) = {1\over \tau}\int_0^\tau dt \, \Delta Q_{a\to b}(t)
\end{align}
plotted in figure \ref{f4}, which shows that the CSL is only violated for very small values of $\tau$.

\begin{figure}[h]
\begin{minipage}{\columnwidth}
\begin{center}
\hspace*{-4mm}\includegraphics[scale=0.7]{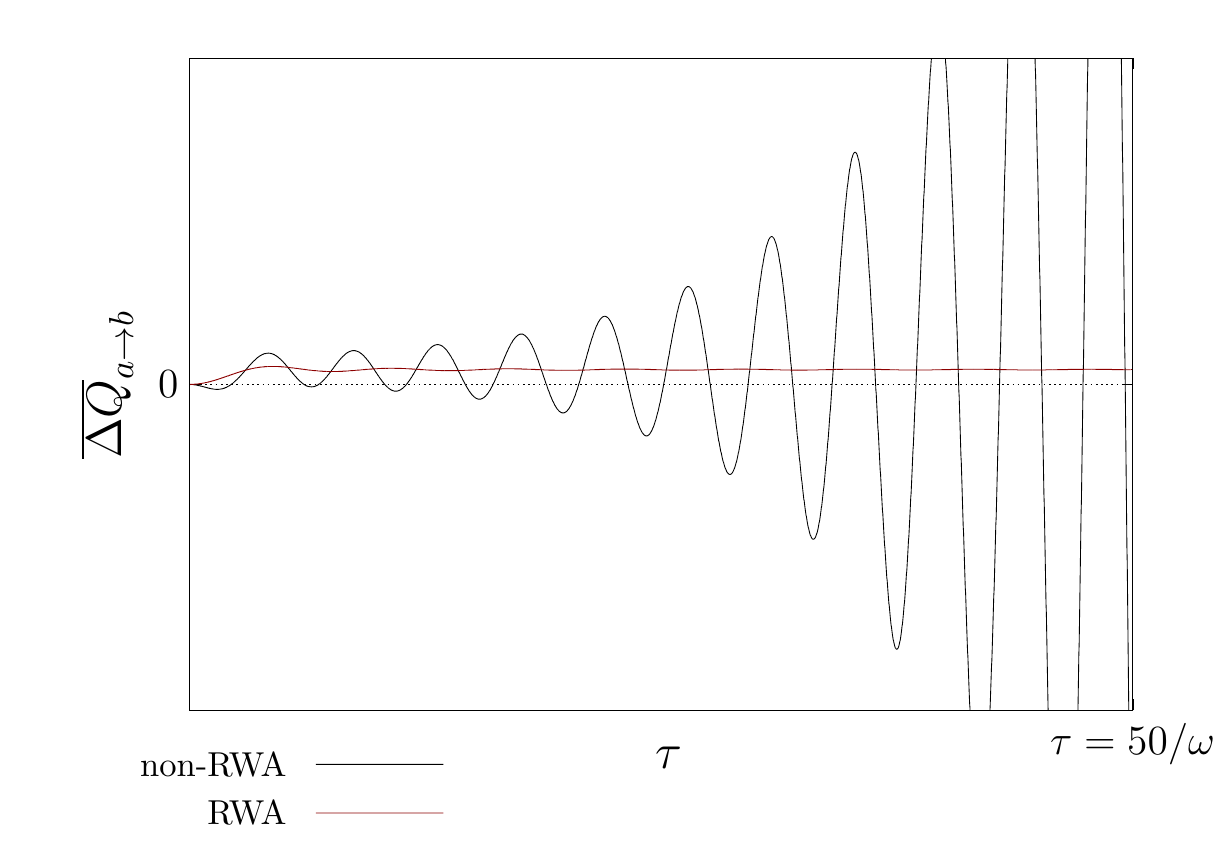}
\end{center}
\vspace*{-5mm}\caption{Plot of $\overline{\Delta Q}_a(\tau)$ in the strong coupling regime with $g/\omega =0.51$ at fixed temperature difference $T_a-T_b =50K$. The time-averged heat transfer exhibits large violations of the CSL for long averaging periods $\tau$. The violations of the CSL cannot be interpreted as transient in this regime.} \label{f5}
\end{minipage}
\end{figure}

The time-averaged heat transfer is ill-defined when $g/\omega = 1/2$, and for large values of $g$, i.e., in the strong coupling regime such that $g/\omega >1/2$, $\overline{\Delta Q}_{a\to b}$ behaves quite differently. The interpretation of the violations as transient is no longer available in this regime as the behavior of $\overline{\Delta Q}_{a\to b}$ indicates extensive violations of the CSL for large $\tau$ (figure \ref{f5}).


\subsection{Discussion}

There are at least two ways in which the violations of the CSL exhibited in section \ref{violatess} can be interpreted. If one assumes the bare energy operators $H_a$ and $H_b$ represent the observable (heat) energies of the subsystems, then the results in section \ref{violatess} show that the CSL is genuinely violated, albeit transiently, within certain parameter regimes and for certain interaction Hamiltonians. In quantum optics such violations might be explained in terms of the energy-time uncertainty principle, which allows for the violation of bare energy conservation, and therefore the spontaneous production of heat, over very short timescales. There is experimental evidence to support such violations \cite{wang_experimental_2002,gieseler_dynamic_2014}. If however, such violations are viewed as untenable, even transiently, then either the physically available interactions between quantum subsystems become limited by the CSL, or one must reject the physicality of the \emph{free} energy operators as representing \emph{observable} energies. With regard to the former of these interpretive options, it is worth remarking that the RWA interaction in (\ref{rwa}) can be obtained as a unitary transformation of the minimal coupling interaction \cite{drummond_unifying_1987,baxter_gauge_1990}. Thus, forms of interaction that necessarily obey the CSL need not be the result of an approximation.

\section{Non-standard definitions of heat, entropy and subsystems}\label{4}

In this section we extend the analysis of the previous sections by considering non-standard definitions of the basic thermodynamic quantities and quantum subsystems.

\subsection{True subsystem energies via non-standard quantum subsystems}\label{true}

The first law of thermodynamics requires the conservation of the total energy of an isolated system. The term \emph{isolated} is preferred in classical thermodynamics, whereas in quantum theory the same kind of system is referred to as \emph{closed}. A closed quantum system is characterised by its unitary dynamics, and the composite $ab$-oscillator system that we are considering is of precisely this type.

Since the entropy in (\ref{S0}) is defined in terms of the free energies $H_a$ and $H_b$ rather than $H$, the first law that corresponds to the ESL (\ref{ESL}) could be taken as the requirement ${\dot H_0}={\dot H}_a+{\dot H}_b = 0$, which is nothing but condition (\ref{result2}). One can therefore understand the violations of the CSL as a consequence of violations of this first law. However, the condition ${\dot H}=0$, which expresses the conservation of the true total energy, and which necessarily holds for a closed quantum system with time-independent Hamiltonian, could also be interpreted as the first law of thermodynamics in the present context.

Our results show that the CSL will hold when the first and second laws are assumed to refer to the same type of energy $H_0$, and the first law is assumed to hold in the form ${\dot H}_0=0$. Alternatively, if one takes ${\dot H}=0$ as expressing the first law, we see that the first and second laws refer to two distinct types of energy whenever there are interactions. In this latter case, it is possible to violate the CSL even though the first law ${\dot H}=0$ necessarily holds, because the sum of subsystem energies $H_0$ does not coincide with the true total energy $H$. In order that $H_a + H_b =H$, we must have either that $V\equiv 0$,
 or we must redefine the subsystem energies $H_a$ and $H_b$.

To investigate the option of redefining the subsystem energies we consider the case of minimally coupled oscillators. Introducing position and momentum variables as
\begin{align}
x_c = \sqrt{1\over 2m\omega} (c^\dagger +c),\qquad p_c = i\sqrt{m\omega \over 2}(c^\dagger-c)
\end{align}
where $c=a,b$ and $m>0$, and minimally coupling oscillator $a$ to $b$ through a coupling parameter $q>0$ gives the Hamiltonian
\begin{align}\label{h}
H={1\over 2m}(p_a - qx_b)^2 +{1\over 2}m\omega^2x_a^2 + {p_b^2\over 2m}+{1\over 2}m\omega^2x_b^2
\end{align}
where for notational simplicity tensor products with the identity operators for $a$ and $b$ have been omitted. Alternatively, one could minimally couple oscillator $b$ to $a$ so as to give the Hamiltonian
\begin{align}\label{h22}
H'={p_a^2\over 2m} +{1\over 2}m\omega^2x_a^2 + {1\over 2m}(p_b + qx_a)^2+{1\over 2}m\omega^2x_b^2.
\end{align}
The two Hamiltonians $H$ and $H'$ are unitarily related by the transformation $U=e^{-iqx_a\otimes x_b}$. As a result they produce identical equations of motion written in terms of the derivatives of $x_a$ and $x_b$ \footnote{Here we are adopting an {\em active} view towards unitary transformations within the composite system's Hilbert space. This means that the Hamiltonian is actively transformed into a new unitarily equivalent Hamiltonian. In the introduction \ref{1} we adopted a {\em passive} viewpoint towards unitary transformations, in which the same Hamiltonian was expressed in different operator bases. The active and passive perspectives are of course equivalent for understanding quantum subsystem relativity.}. However the non-local (in the sense of tensor product structure) transformation $U$, which does not commute with $p_a$ and $p_b$ results in these operators taking on different physical meanings with respect to the distinct Hilbert space representations connected by $U$. Because of this the free energies $H_a$ and $H_b$ are also physically different within the two different Hilbert space representations. We see that conventional subsystem observables are highly non-unique and therefore physically ambiguous as noted in Section \ref{1}.

One can define unique subsystem observables in terms of $x_a$, $x_b$ and their velocities. In particular both $H$ and $H'$ can be written in the invariant form
\begin{align}\label{ht2}
H={1\over 2}m({\dot x}_a^2+\omega^2x_a^2) + {1\over 2}m({\dot x}_b^2+\omega^2x_b^2) = H'
\end{align}
showing that they are in fact identical Hamiltonians, with alternative forms in (\ref{h}) and (\ref{h22}) that result from their superficial expression in terms of physically distinct canonical momenta. It is clear from (\ref{ht2}) that one can define unique and physically unambiguous subsystem energies $H_a^{\rm true}$ and $H_b^{\rm true}$ such that 
\begin{align}\label{habtr}
H_c^{\rm true} = {1\over 2}m({\dot x}_c^2+\omega^2x_c^2)
\end{align}
with $c=a,b$. These definitions are especially natural in that they express the total energy $H$ as the sum of subsystem energies. The subsystems are moreover defined in terms of precisely the same physical variables as in the free (non-interacting) case, namely the positions $x_a$ and $x_b$, and the \emph{mechanical} momenta $m{\dot x}_a$ and $m{\dot x}_b$.

It is important to recognise that this definition of a quantum subsystem cannot coincide with the conventional one, because ${\dot x}_a$ and ${\dot x}_b$ do not commute. This means that there are no operators ${\dot {\tilde x}}_a$ and ${\dot {\tilde x}}_b$ such that ${\dot x}_a= {\dot {\tilde x}}_a \otimes I_b$ and \emph{simultaneously} ${\dot x}_b = I_a\otimes {\dot {\tilde x}}_b$. As such the physical subsystems are not determined through a (physically non-unique) mathematical decomposition of the Hilbert space based on tensor product structure, but rather in terms of their assumed to be fundamental dynamical variables.

With this definition of a quantum subsystem the interaction between $a$ and $b$ is described by the non-commutativity of the velocity operators, rather than through an interaction Hamiltonian $V$, and it is not possible to simultaneously prepare the subsystems in states of well-defined energy, or well-defined temperature as in (\ref{rho02}). One cannot then pose sharp questions about heat transfer between the two systems and the apparent violations of the CSL in section \ref{violatess} lose any physical significance.

It is natural to ask whether there are, from the new point of view, \emph{any} conditions under which questions about heat transfer can be meaningfully posed. By way of an answer, note that according to (\ref{h}) $H_b^{\rm true}=H_b$ while $H_a^{\rm true}=H-H_b$, and yet according to (\ref{h22}) $H_a^{\rm true}=H_a$ while $H_b^{\rm true} = H'-H_a$. Since performing the RWA and neglecting the self-energy terms in either $H$ or $H'$ produces the same resulting Hamiltonian (when the oscillators are resonant), consistency requires that within the regime in which these approximations are justified we must have that $H_a^{\rm true} \approx H_a$ \emph{and} $H_b^{\rm true} \approx H_b$. Thus, the rotating-wave form of the Hamiltonian is the unique form in which \emph{both} of the bare energy operators $H_{a,b}$ approximately coincide with the corresponding true energies.

From the current perspective, the initial state in (\ref{rho02}) cannot truly specify the individual states of the subsystems, because the true subsystems do \emph{not} specify a tensor product decomposition of the composite Hilbert space ${\mathcal H}={\mathcal H}_a\otimes {\mathcal H}_b$ such that the factors ${\mathcal H}_{a,b}$ represent the \emph{physical} subsystem state spaces. At best the initial state in (\ref{rho02}) can be considered an approximation, and the analysis above demonstrates that the conditions under which this approximation is valid are precisely those conditions under which the CSL cannot be violated. We can therefore explain the apparent violations of the CSL as artifacts of an unwarranted attribution of sharp temperatures to the subsystems, when such attributions are not strictly permissible, and can hold only approximately. This explanation is similar in nature to that offered by the energy-time uncertainty relation, which suggests that there is always some uncertainty in the (heat) energies of the subsystems, so that on short enough timescales apparent violations of the CSL may occur.

\subsection{Entropy production and heat exchange}

In \cite{esposito_entropy_2010,esposito_second_2011} Esposito \emph{et al.}~investigate the production of entropy within a system coupled to several thermal reservoirs in thermal equilibrium. Since no constraints are placed on what actually constitutes a reservoir, the treatment there is easily adapted to the situation considered here such that each oscillator is considered as a reservoir for the other. The authors in \cite{esposito_entropy_2010,esposito_second_2011} show that the total change in the von Neumann entropy of system $a$ can be written assuming an initially separable state
\begin{align}
\Delta S_a(t) &= -{\rm tr}[\rho_a(t)\ln\rho_a(t)] + {\rm tr}[\rho_a\ln\rho_a] \nonumber \\ &= \Delta_iS_a(t) + \Delta_eS_a(t)
\end{align}
where $\Delta_i S_a(t)$ denotes the irreversible entropy production within system $a$ and $\Delta_e S_a(t)$ denotes the reversible entropy-flux due to the reservoir system $b$. Explicitly these contributions are given by
\begin{align}\label{dide}
\Delta_i S_a(t) &=S(\rho(t)||\rho_a(t)\otimes\rho_b)\nonumber \\
\Delta_e S_a(t) & = \beta_b(\langle H_b(0)\rangle -\langle H_b(t)\rangle ) = -\beta_b \Delta Q_b(t)
\end{align}
where $S(\rho||\sigma) :={\rm tr}(\rho\ln\rho)-{\rm tr}(\rho\ln\sigma)$ denotes the relative entropy between states $\rho$ and $\sigma$, and where $\beta_b \Delta Q_b(t)$ denotes the contribution from system $b$ to the free entropy change given in (\ref{S0}). 

One possibility at this point is to use the Gibbs relation (\ref{gibbs}) and the identifications made in (\ref{dide}) to define the heat flux's into systems $a$ and $b$ by
\begin{align}\label{flx2}
{\tilde \Delta Q}_a(t) &= {1\over \beta_a}\Delta_e S_a(t) =-{\beta_b\over \beta_a}\Delta Q_b(t),\nonumber \\
{\tilde \Delta Q}_b(t) &={1\over \beta_b}\Delta_e S_b(t)=  -{\beta_a\over \beta_b}\Delta Q_a(t).
\end{align} 
These definitions are essentially the reverse of those given in section \ref{theory}. As such the proof of the CSL given these definitions follows almost exactly as in \ref{theory}. Assuming ${\dot H}_0 = 0$ we obtain the CSL in the form
\begin{align}
(\beta_a-\beta_b){\tilde \Delta Q}_{a\to b} = -\left[{\beta_a\over \beta_b}+ {\beta_b\over \beta_a}\right]\Delta S_0 \leq 0.
\end{align}
Here, as in section \ref{theory}, when $[H,V]\neq 0$ the CSL does not follow from the ESL $\Delta S_0 \geq 0$, and the results in section \ref{violatess} remain valid.

Since the above identification (\ref{flx2}) uses the Gibbs relation as in section \ref{theory} it is not suprising that the proof of the CSL requires the same additional assumption (\ref{result2}). However, in \cite{esposito_entropy_2010,esposito_second_2011} the connection between the heat flux defined in terms of $\Delta_e S_a(t)$ and the first law of thermodynamics is discussed without ever involving the Gibbs relation. If we consider system $b$ as a reservoir for $a$ then the heat flux from the reservoir into $a$ can be identified as $-\Delta Q_b(t)$. The authors in \cite{esposito_entropy_2010,esposito_second_2011} use this definition and the fact that ${\rm tr}[H(t){\dot \rho}(t)]=0$ to identify the heat and work contributions to the energy change within system $a$ as follows
\begin{align}\label{es}
\langle \Delta H^{\rm true}_a\rangle &= \langle \Delta Q_a \rangle + \langle \Delta W_a \rangle, \nonumber  \\
\langle \Delta Q^{\rm true}_a \rangle &= \int_0^t dt' \,{\rm tr}[(H_a(t)+V(t)){\dot \rho}_a(t)],\nonumber \\
\langle \Delta W^{\rm true}_a \rangle &= \int_0^t dt' \,{\rm tr}[({\dot H}_a(t)+{\dot V}(t))\rho_a(t)].
\end{align}
In our situation $H$ is not explicitly time-dependent in the Schr\"odinger picture so the work contribution vanishes and
\begin{align}\label{nonst}
\langle \Delta H^{\rm true}_a\rangle &= \langle \Delta Q^{\rm true}_a \rangle \nonumber \\ &= {\rm tr}[(H_a+V)\rho_a(t)]-{\rm tr}[(H_a+V)\rho_a].
\end{align}
According to (\ref{nonst}) the physical energy of system $a$ is an observable $H^{\rm true}_a=H_a+V=H-H_b$ as in section \ref{true}. This true energy is defined in terms of operators that according to the standard definition of a quantum subsystem are associated with system $b$. Therefore, according to this definition the true subsystems $a$ and $b$ are not standard quantum subsystems. Furthermore, since we are considering a situation in which the labels ``system" and ``reservoir" can be interchanged, we see that naively applying the same method as above in order to obtain the energy-flux into system $b$ from system $a$ gives
\begin{align}\label{nonst2}
\langle \Delta Q^{\rm true}_b \rangle = {\rm tr}[(H_b+V)\rho_b(t)]-{\rm tr}[(H_b+V)\rho_b].
\end{align}
Now, if the energy lost by system $a$ ($b$) must equal that gained by system $b$ ($a$) then (\ref{nonst2}) cannot generally be reconciled with (\ref{dide}) and (\ref{nonst}) using only a single tensor-product decomposition of the Hilbert space into subsystem state spaces. This means one must either identify subsystems in a non-standard way using more than one tensor product decomposition, or allow the non-conservation of the sum of subsystem energies. The latter option means allowing $[H,V]\neq 0$, provided of course, that we use the standard definition of subsystem energies. If, instead, one uses within one and the same tensor-product decomposition, the non-standard definition where
\begin{align}\label{nst}
H_a^{\rm true} = H-H_b \, , ~~~ H_b^{\rm true} = H-H_a
\end{align}
as supplied in \cite{esposito_entropy_2010,esposito_second_2011} and used in (\ref{nonst}) and (\ref{nonst2}), then one obtains $H_a^{\rm true}+H_b^{\rm true} = H+V$, which is not generally a conserved quantity. Moreover, adopting (\ref{nst}) the heat-flux between the systems actually remains unchanged;
\begin{align}\label{o}
\Delta Q_{a \to b}^{\rm true} &:= \langle H_a^{\rm true}(t)\rangle - \langle H_b^{\rm true} \rangle - \langle H_b^{\rm true}(t) \rangle + \langle H_b^{\rm true} \rangle \nonumber \\ &\equiv \Delta Q_{a\to b},
\end{align}
and we know that the CSL for the quantity $\Delta Q_{a\to b}$ will not generally hold unless condition (\ref{result2}) is met.

As with the energies $H_{a,b}$, the sum of the alternative energies in (\ref{nst}) will only coincide with the true total energy when $V=0$. Thus, as explained in section \ref{true}, one might argue that within a single tensor-product decomposition the two equalities in (\ref{nst}) can only approximately hold simultaneously when the coupling is sufficiently weak; $V\approx 0$. The CSL can then be recovered within the regime for which using (\ref{nst}) in conjunction with a single tensor-product decomposition constitutes a valid approximation.

In summary then, if we adopt as in \cite{esposito_entropy_2010,esposito_second_2011}, the alternative identifications of entropy and heat given by (\ref{dide}) and (\ref{es}) we reach essentially the same conclusions as before; either we accept that the CSL is violated, or we accept the restrictions it places upon the form of the interaction, or we adopt a non-standard definition of quantum subsystem. It is this last option that appears to have been implicitly assumed in \cite{esposito_entropy_2010}.

\subsection{The CSL with non-standard subsystem energies}

The identification of the true heat flux made in (\ref{o}) \cite{esposito_entropy_2010,esposito_second_2011}, is consistent with the analysis made in section \ref{true} whereby it was argued that the initial thermal state in (\ref{rho0}) cannot be justified when (\ref{result2}) does not hold. It is obviously justified however, if at $t=0$ the interaction is adiabatically switched on. This gives a time-dependent Hamiltonian $H(t) = H_0 + V(t)$ such that $H(0)=H_0$. Since the Hamiltonian is time-dependent the work contribution to the expected energy no longer vanishes.

We now adopt the non-standard definition of subsystem energies used in section \ref{true}, and assume without loss of generality that work is performed on system $a$. We can visualise this situation through some mechanism that gradually introduces system $a$ to system $b$, which remains stationary. Such a situation could be modelled in the case of minimally coupled oscillators by using the Hamiltonian in (\ref{h}) with the coupling constant $q$ replaced by a time-dependent function $q(t)$. The energy operators for the oscillator subsystems then become
\begin{align}
H_a^{\rm true}(t) = {1\over 2}m({\dot x}_a(t)^2 + x_a^2),\qquad H_b = {1\over 2}m({\dot x}_b^2 + x_b^2)
\end{align}
with $m{\dot x}_a(t) = p_a -q(t)x_b$. If instead we were to use the Hamiltonian in (\ref{h22}) we would be describing a situation in which work were being performed on system $b$ with
\begin{align}
H_b^{\rm true}(t) = {1\over 2}m({\dot x}_b(t)^2 + x_b^2), \qquad H_a = {1\over 2}m({\dot x}_a^2 + x_a^2)
\end{align}
and $m{\dot x}_b(t) = p_b+q(t)x_a$. We see that when the energy is explicitly time-dependent the two Hamiltonians in (\ref{h}) and (\ref{h22}) describe distinct situations. 

Since work is being performed on the system we should no longer expect that heat cannot be displaced against a temperature gradient. Assuming that work is performed on system $a$ we obtain from ${\rm tr}(H{\dot \rho})=0$ that
\begin{align}\label{oo}
\langle \Delta Q^{\rm true}_a \rangle = \int_0^t dt' {\rm tr}[H_a^{\rm true}(t'){\dot \rho}(t')] = -\langle \Delta Q_b \rangle.
\end{align}
If we now define
\begin{align}
\Delta S_Q = \beta_a\langle \Delta Q_a^{\rm true}\rangle + \beta_b\langle \Delta Q_b\rangle 
\end{align}
we immediately obtain
\begin{align}
\Delta S_Q \geq 0 \Leftrightarrow (\beta_b -\beta_a)\langle \Delta Q_{a\to b}^{\rm true}\rangle\geq 0.
\end{align}
so that with non-standard identifications of subsystem energies the inequality for heat entropy increase is strictly equivalent to the Clausius inequality constraining heat transfer. Of course, unlike the situation with $\Delta S_0$ we no longer have that $\Delta S_Q \geq 0$ in general. Interestingly, however, the condition $[H_0,V(t')]$, which is a time-dependent generalisation of condition (\ref{result2}), can now be seen as the condition required in order that the above definitions  reduce to the standard definitions made in section \ref{standard}. To see this we write the heat change in (\ref{oo}) as
\begin{align}
\langle \Delta Q^{\rm true}_a\rangle = \langle\Delta Q_a \rangle+ i\int_0^t dt' {\rm tr}([H_0,V(t')]\rho(t'))
\end{align}
so that
\begin{align}
[H_0,V(t')]&=0 \Rightarrow \nonumber \\ &\langle \Delta Q^{\rm true}_a\rangle = \langle\Delta Q_a \rangle~~ {\rm and} ~~\Delta S_Q = \Delta S_0.
\end{align}
Since $\Delta S_0 \geq 0$ (the proof in section \ref{con} continues to hold for time-dependent Hamiltonians) we see that $[H_0,V(t')]=0$ is the condition for which the system continues to obey the Clausius relation $(\beta_b-\beta_a)\langle \Delta Q^{\rm true}_{a\to b} \rangle\geq 0$ despite being externally driven. This gives a criterion by which it would be impossible to create a microscopic fridge by interacting two quantum systems when assuming the non-standard definition of subsystem energies (cf.~Section \ref{true}).

\subsection{Heat exchange in the energy-flux formalism}

In \cite{weimer_local_2008}, alternative definitions to those given by (\ref{heat}) are given for the heat and work energies of interacting subsystems. The formalism introduced there has certain advantages and is further extended in \cite{hossein-nejad_work_2015}. It is of interest to relate the standard definitions given in section \ref{theory} to the definitions given in \cite{weimer_local_2008}, in the case of the interacting oscillator system currently under consideration. For the remainder of this section we work in the Schr\"odinger picture, but allow for the possibility of explicitly time-dependent observables $O(t)$ such that ${\dot O}\neq 0$

In \cite{weimer_local_2008} it is assumed that a so-called local measurement basis (LEMBAS) for system $a$ is determined via coupling to a measurement device. The LEMBAS chosen in \cite{weimer_local_2008} is the bare energy basis associated with $H_a$. An effective Hamiltonian is defined in the Schr\"odinger picture via
\begin{align}\label{heff}
H_a^{\rm eff}(t) := {\rm tr}_b[V I_a\otimes \rho_b(t)],
\end{align}
where $\rho_b(t)={\rm tr}_a\rho(t)$. The effective Hamiltonian can be partitioned as $H_a^{\rm eff} = H_1^{\rm eff}+H_2^{\rm eff}$ where $H_1^{\rm eff}$ is the component of $H_a^{\rm eff}$ that is diagonal in the eigenbasis of $H_a$ and which therefore commutes with $H_a$. The component $H_2^{\rm eff}$ does not commute with $H_a$ unless $H_2^{\rm eff} \equiv 0$. The total energy of system $a$ is assumed to be represented by a Hamiltonian $H'_a$ defined as
\begin{align}
H_a'=H_a+H_1^{\rm eff}.
\end{align}
This is motivated by the idea that only $H_1^{\rm eff}$ will contribute in a measurement for which the LEMBAS is the eigenbasis of $H_a$. The heat and work components within the relation (\ref{1st}) can be identified within this framework by comparison with the dynamics of subsystem $a$ which are given in the Schr\"odinger picture by
\begin{align}\label{rhodot}
{\dot \rho}_a = -i{\rm tr}_b([H,\rho]) = -i[H_a +H_a^{\rm eff},\rho_a] + {\rm tr}_b{\mathcal L}(\rho)
\end{align}
where $\rho_a(t) :={\rm tr}_b\rho(t)$ and ${\mathcal L}$ is a super-operator that acts on the composite state. The presence of the second term ${\rm tr}_b{\mathcal L}(\rho)$ in (\ref{rhodot}) indicates that the reduced dynamics of system $a$ are non-unitary. The work and heat contributions to the rate of change of the total energy are then given as contributions arising from the unitary and non-unitary components of (\ref{rhodot}) respectively;
\begin{align}\label{new}
&{d\langle H_a' \rangle \over dt} = {\rm tr}({\dot H}_a'\rho_a) + {\rm tr}(H_a' {\dot \rho}_a),\nonumber \\
&{d\langle W_a\rangle \over dt} = {\rm tr}({\dot H}'_a\rho_a-i[H_a',H_2^{\rm eff}]\rho_a), \nonumber \\ &{d\langle Q_a \rangle \over dt}= {\rm tr}{\mathcal L}(\rho).
\end{align}
In \cite{weimer_local_2008} it is further assumed that ${\dot H}_a=0$ (in the Schr\"odinger picture) so that ${\dot H}_a' = {\dot H}_1^{\rm eff}$.

In contrast to the definitions given by (\ref{heat}), in (\ref{new}) the work and heat contributions do not coincide with the individual terms on the right hand side of the first equality. However, the two sets of definitions are essentially the same whenever $H_2^{\rm eff}=0$. In this case the energy of system $a$ is represented by a Hamiltonian $H_a'$ which is identical to $H_a$, but with the bare frequencies replaced by renormalised frequencies $\omega_a^{n,{\rm ren}} = \omega_a^n+\bra{n_a}H_a^{\rm eff}\ket{n_a}$. This situation would occur if one were to assume that the LEMBAS were given as the eigenbasis of $H_a+H_a^{\rm eff}$. This is perhaps a natural choice given that the physical energy of system $a$ is supposed to be represented by an operator $H_a' \neq H_a$, so that $H_a$ loses any particular physical significance.

As is pointed out in \cite{weimer_local_2008} the LEMBAS framework does not provide a means by which to identify the LEMBAS itself. The LEMBAS is supposed to be determined through a detailed analysis of the interaction with the measuring device. In particular, there is no obvious justification for assuming that a measuring device selects the eigenbasis of the apparently non-physical energy $H_a$ as being preferred.

In general it appears that whether or not the CSL is violated may depend on the LEMBAS chosen, which could therefore be taken as providing an additional criterion by which a LEMBAS can be identified. As pointed out above if we adopt the obvious choice of the eigenbasis of $H_a'$ then the definitions of heat and work are essentially the same as the standard ones and the results of the previous sections continue to hold.

If, as in \cite{weimer_local_2008} we choose the eigenbasis of $H_a$ as the LEMBAS and we again consider the interacting oscillator system considered in section \ref{vio}, then due to the linearity of the solutions (\ref{abgen}), we find that $H_a^{\rm eff}$ vanishes. To see this we write the Schr\"odinger picture interaction Hamiltonian in (\ref{v}) as
\begin{align}
V= ig(a^\dagger + a)\otimes (b^\dagger-b) \equiv V_a\otimes V_b
\end{align}
and we express the initial states $\rho_a$ and $\rho_b$ in (\ref{rho0}) through their spectral representations which equal
\begin{align}\label{spec}
\rho_c = \sum_n \lambda_c^n \ket{n_c}\bra{n_c},\qquad \lambda_c^n :={e^{-\beta_c n_c \omega_c}\over Z_c} 
\end{align}
where $c=a,b$. We then obtain
\begin{align}\label{z}
H_a^{\rm eff} =& {\rm tr}_b\left(V_a\otimes V_b \, \,I_a\otimes {\rm tr}_a[U(t)\rho(0) U^\dagger(t)]\right) \nonumber \\ =&V_a{\rm tr}\left(\sum_{nm} \lambda_a^n\bra{n_a}U^\dagger(t)\ket{m_a}V_b\bra{m_a}U(t)\ket{n_a}\rho_b\right)\nonumber \\
=&V_a{\rm tr}\left(\sum_n \lambda_a^n \bra{n_a}U^\dagger(t)(I_a\otimes V_b)U(t)\ket{n_a}\rho_b\right)
\end{align}
where the completeness of the bare states $\{\ket{n_a}\}$ and the cyclicity of the trace have been used. Since $V_b(t)=U^\dagger(t)I_a\otimes V_bU(t)$ is linear in the operators $a,a^\dagger, b, b^\dagger$, and $\rho_b$ in (\ref{spec}) is diagonal, the trace term on the last line of (\ref{z}) vanishes. This shows that choosing the eigenbasis of $H_a$ as the LEMBAS, the energy-flux framework coincides with the standard framework (section \ref{standard}), and the results of section \ref{violatess} continue to apply. Thus, neither the eigenbasis of $H'_a$ nor that of $H_a$, which are both obvious choices of LEMBAS, result in an avoidance of the violations of the CSL seen in section \ref{violatess}.
 
\section{Conclusions}\label{conc}

In this paper we use classical thermodynamics to eliminate ambiguity in quantum physics. More concretely, we introduce a thermodynamic constraint to identify the possible subsystem decompositions of a composite quantum system. In general, there are many ways to write the total Hilbert space $\cal H$ of a composite quantum system as a tensor product of subspaces. Different subsystem definitions generally imply different interaction Hamiltonians and different expectation values for subsystem observables. The identification of physical subsystems is therefore of fundamental importance for ensuring the uniqueness of physical predictions \cite{stokes_extending_2012,stokes_gauge_2014}. 

In section \ref{theory}, we use the standard definitions of thermodynamic quantities to identify a sufficient condition which ensures that the heat transfer between two quantum subsystems obeys Clausius' form of the second law of thermodynamics. It is shown that this law holds, whenever the unique total Hamiltonian $H$ and the subsystem-dependent interaction Hamiltonian $V$ commute (cf.~(\ref{result2})).
Section \ref{vio} shows that violations of Clausius' law can occur when this condition is not met. Our results imply either the possibility of genuine violations of the CSL, or the physical impossibility of interactions $V$ that do not commute with the total Hamiltonian $H$, and hence the physical impossibility of their associated subsystem decompositions. This provides some resolution to an ongoing debate on how to identify appropriate interaction Hamiltonians in quantum physics (cf.~ \cite{power_coulomb_1959,power_nature_1978,drummond_unifying_1987,cohen-tannoudji_photons_1989,baxter_gauge_1990,woolley_gauge_2000,stokes_extending_2012,stokes_gauge_2013}). Finally, Section \ref{4} extends our analysis to non-standard definitions of heat, work and entropy that have recently been proposed in the literature \cite{weimer_local_2008,esposito_entropy_2010,esposito_second_2011,hossein-nejad_work_2015}. It is shown that none of these alternative identifications modify the conclusions drawn from the standard analysis in section \ref{vio}. \\[0.5cm]
{\em Acknowledgments.} AS and AB acknowledge financial support from the UK Engineering and Physical Sciences Research Council EPSRC (Grant Refs.~EP/H048901/1 and EP/M013243/1). PD thanks the British Council and DST Government of India for financial support.

\bibliography{Thermo.bib}

\end{document}